\documentclass{article}

\usepackage{arxiv}
\usepackage[numbers]{natbib}

\usepackage{amsmath, enumitem}
\usepackage{amsfonts}
\usepackage{upgreek}
\usepackage{graphicx}
\graphicspath{{Figures/}} 
\usepackage{setspace}
\usepackage{multirow}
\usepackage[T1]{fontenc}
\usepackage{fixltx2e}
\usepackage{xcolor}
\usepackage[skins,breakable]{tcolorbox}

\newtcolorbox{cvbox}[2][]{%
  blanker,
  width = 0.9\textwidth,
  after skip=8mm,
  title=#2,
  breakable,
  #1
}

\title{Surface wave dispersion inversion using an energy likelihood function}

\author{
	Xin Zhang \\
	School of Geosciences \\
	University of Edinburgh\\
	Edinburgh, United Kingdom \\
	\texttt{x.zhang2@ed.ac.uk} \\
	\And
	York Zheng \\
	BP p.l.c. \\
    London, United Kingdom \\
	\And
	Andrew Curtis \\
	School of Geosciences \\
	University of Edinburgh\\
	Edinburgh, Unite Kingdom \\
	\texttt{andrew.curtis@ed.ac.uk} \\
}
\begin{document}
	\maketitle

\begin{abstract}
Seismic surface wave dispersion inversion is used widely to study the subsurface structure of the Earth. The dispersion property is usually measured by using frequency-phase velocity (f-c) analysis and by picking phase velocities from the obtained f-c spectrum. However, because of potential contamination the f-c spectrum often has multimodalities at each frequency for each mode. These introduce uncertainty and errors in the picked phase velocities, and consequently the obtained shear velocity structure is biased. To overcome this issue, in this study we introduce a new method which directly uses the spectrum as data. We achieve this by solving the inverse problem in a Bayesian framework and define a new likelihood function, the energy likelihood function, which uses the spectrum energy to define data fit. We apply the new method to a land dataset recorded by a dense receiver array, and compare the results to those obtained using the traditional method. The results show that the new method produces more accurate results since they better match independent data from refraction tomography. This real-data application also shows that it can be applied efficiently since it removes the need to pick phase velocities, and with relatively little adjustment to current practice since it uses standard f-c panels to define the likelihood. We therefore recommend using the energy likelihood function rather than explicitly picking phase velocities in surface wave dispersion inversion.
\end{abstract}

\section{Introduction}
Seismic surface waves travel along the surface of the Earth while oscillating over depth ranges that depend on their frequency of oscillation \citep{aki1980quantitative}. This in turn makes surface waves dispersive -- different frequencies travel at different speeds, and these speeds are sensitive to different parts of the Earth. By measuring speeds at different frequencies this dispersion property can be used to constrain subsurface structures over different depth ranges on global scale \citep{trampert1995global, shapiro2002monte, meier2007aglobal, meier2007bglobal}, regional scale \citep{zielhuis1994deep, curtis1998eurasian, simons2002multimode, yao2006surface} and industrial scale \citep{park1999multichannel, xia2003inversion, de2011ambient, zhang20201d}.

Surface wave dispersion property (phase or group velocities at different frequencies) can be measured in different ways depending on different acquisition systems. In the case of a single station or a sparse receiver array, as is often the case in seismology, the dispersion property can be measured by using the frequency-time analysis (FTAN) method \citep{dziewonski1969technique, levshin1972frequency, levshin1992peculiarities, herrin1977phase, russell1988application, ritzwoller1998eurasian, levshin2001automated, bensen2007processing}. In FTAN one constructs a frequency-time domain envelope image for each seismic trace by using a set of narrow bandpass Gaussian filters, and measures the group velocity using the arrival time of the maximum envelope at each frequency. The phase velocity can be derived using the phase of the signal at the time of the maximum envelope plus a phase ambiguity term (appropriate integer multiple of $2\pi$) and a source phase term, or by using an image transformation technique \citep{yao2006surface}. However, the obtained measurements are often uncertain because of the trade-off between the determination of frequency of a signal and the determination of arrival time of a signal, and because of the contamination contained in the envelope image caused by effects such as multipathing. In addition, those methods cannot discriminate between different modes, which introduces errors to the measurements made for the fundamental or other modes and cannot provide measurements for higher modes.

In the case where a dense receiver array is deployed, the phase velocity can be determined using frequency-phase velocity (f-c) analysis, in which wavefields recorded by the array are slant stacked to obtain a f-c spectrum \citep{park1998imaging, xia2003inversion}. The phase velocity is then determined as the velocity associated with the highest energy at each frequency in the spectrum. Because different modes are separated in the f-c domain, the method can also be applied to determine phase velocities for higher modes. However, the obtained spectrum may still be contaminated by multipathing effects or strong lateral heterogeneity \citep{hou2016multi}, and consequently the velocity associated with the highest energy does not necessarily represent the most appropriate phase velocity measurement. This issue can be overcome by imposing additional prior information to the phase velocity, for example, by forcing the picked phase velocity dispersion curve to be continuous. Unfortunately, on one hand such additional constrains are usually achieved by picking the phase velocity deliberately and manually for each spectrum, which cannot be applied to large datasets. On the other hand an automatic procedure can easily introduce errors to the picked phase velocities because of the complexity of the spectrum.

To overcome these issues in phase velocity estimation, in this study we introduce a new method which directly uses the spectrum as data rather than explicit picks of the phase velocities. The spectrum of data has been used in wave-equation dispersion inversion in the framework of full-waveform inversion \citep{li2017wave}. However, that method is computationally expensive and the problem is solved using a deterministic method which cannot provide uncertainty estimates. To quantify uncertainty we solve the dispersion inversion problem using Bayesian inference. In Bayesian inference one constructs a so-called posterior probability density function (pdf) that describes the remaining uncertainty of models post inversion, by combining prior information with the new information contained in the data as represented by a pdf called the likelihood. The likelihood function describes the probability of observing data given a specific model, and is traditionally assumed to be a Gaussian distribution centred on the picked phase velocities. In this study we propose a new likelihood function, called the \textit{energy} likelihood function which directly uses the spectrum based on the intuition that higher energy in the spectrum reflects higher probability of observing the associated phase velocity. 

We apply the new method to a land dataset recorded by a dense array and compare the results with those obtained using the traditional method. The dataset consists of raw shot records taken from a sub-area of a nodal land seismic survey that was conducted in a desert environment \citep{ourabah2020184}.  This dataset offers ultra-high trace density with over 180 million traces per km\textsuperscript{2} on a 12.5 m $\times$ 12.5 m receiver grid and a 100 m $\times$12.5 m source grid.  The increased trace density greatly improves the spatial sampling of the wavefield, which in turn benefits the recording and analysis of surface waves.  As a part of the depth model building process, refraction tomography was performed to yield a shallow P-wave velocity model \citep{buriola2021ultra}, which is then used here for qualitative comparison with the shallow S-wave velocity model obtained from our method.

To solve the Bayesian inference problem, we use the reversible-jump Markov chain Monte Carlo (rj-McMC) method. The rj-McMC method is a generalized McMC method which allows a trans-dimensional inversion to be carried out, meaning that the dimensionality of parameter space (the number of parameters) can vary in the inversion \citep{green1995reversible}. Thus the parameterization itself can be dynamically adapted to the data and to the prior information. The method has been used to estimate phase or group velocity maps of the crustal structure \citep{bodin2009seismic, galetti2015uncertainty, zheng2017transdimensional} and to estimate shear velocity structures of the crust and upper mantle using surface wave dispersion data \citep{bodin2012transdimensional, shen2012joint, young2013transdimensional, galetti2017transdimensional, zhang20201d}.

In the following section we first perform frequency-phase velocity analysis for the recorded data to obtain the f-c spectrum around each geographic location. In section 3 we introduce the new energy likelihood function and give an overview of the rj-McMC algorithm. We then apply the new likelihood function to the obtained spectra to estimate the shear velocity structure, and compare the results with those obtained using the traditional method. The results demonstrate that the new method can generate more accurate results than the traditional method, and can be applied efficiently to large data sets. We therefore conclude that the energy likelihood function provides a valuable tool for surface wave dispersion inversion.

\section{Surface wave dispersion analysis}

\begin{figure}
	\centering
	\includegraphics[width=1.\linewidth]{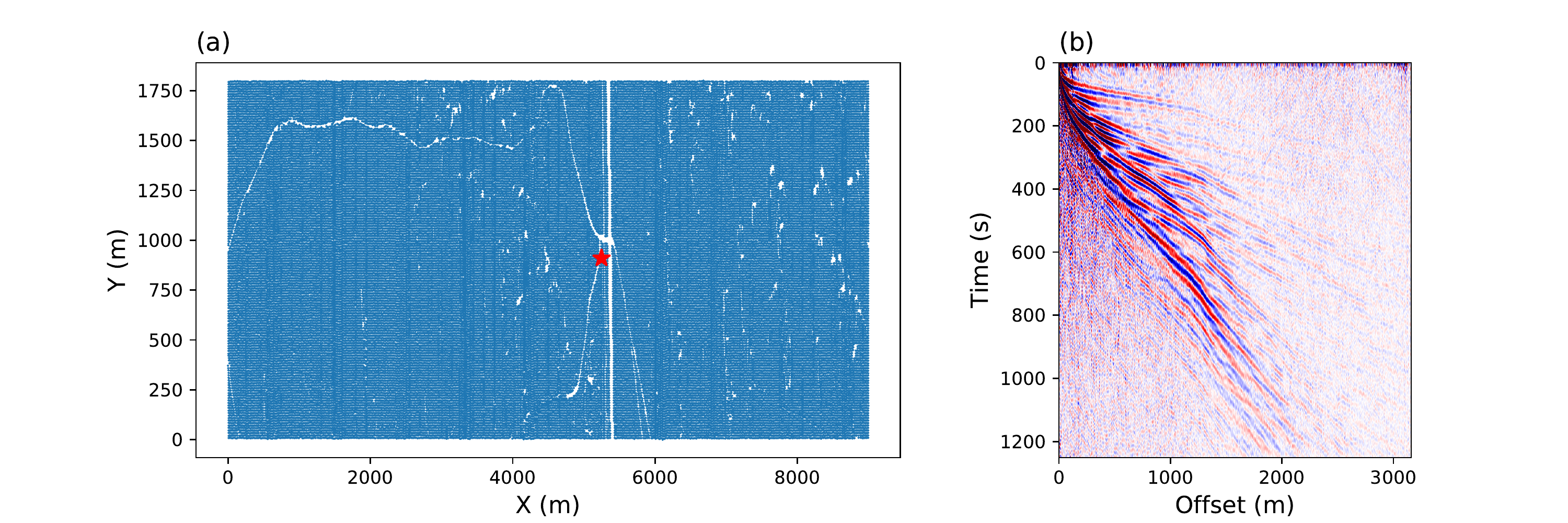}
	\caption{\textbf{(a)} The acquisition system. Blue dots denote receiver locations. The red star is referred to in the text. \textbf{(b)} shows an example of a shot gather, which displays the waveform data generated by a specific shot and recorded by receivers at different offsets.}
	\label{fig:data}
\end{figure}

Figure \ref{fig:data}a shows the locations of all 100,627 sensors which are deployed in a regular grid with a spacing of approximately 12.5 m in both directions, and record samples at 250 Hz. In total, 70261 active sources are fired with 100 m spacing in X direction and 12.5 m spacing in Y direction to generate seismic surface waves. Figure \ref{fig:data}b shows an example of a shot gather which mainly contains surface waves.

To analyse the surface wave dispersion, we performed frequency-phase velocity (f-c) analysis of the recorded data. For a given geographic location $p$, the f-c spectrum $U_{p}(c,w)$ can be computed using the data recorded by a receiver array around the location:
\begin{equation}
 \begin{aligned}
	U_{p}(c,w) &= \int_{\mathcal{C}_{p}} e^{-j\frac{w}{c}x} u(x,w)/|u(x,w)| dx \\
			  &\approx \sum_{i=1}^{N_{p}} e^{-j\frac{w}{c}x_{i}} u(x_{i},w)/|u(x_{i},w)| \Delta x	
 \end{aligned}
 \label{eq:slantstack}
\end{equation}
where $\mathcal{C}_{p}$ denotes that the integration is performed around the location $p$, $x$ is the source-receiver distance, $w$ is frequency in radian, $c$ is phase velocity, $j=\sqrt{-1}$, $i$ is the index of records and $N_{p}$ is the number of receivers around location $p$; $u(x,w)$ is the Fourier transform of the wavefield $u(x,t)$. For a given receiver array a larger $N_{p}$ improves resolution of phase velocity, but reduces spatial resolution. In this study for a given location $p$ we stack all the records whose receiver and source locations are respectively within 300 m and 1500 m to the location $p$. These threshold distances are selected such that the phase velocity dispersion curve can be clearly identified in the spectrum without increasing the number of records unnecessarily. This process is repeated for every geographic location on a regular grid with a 12.5 m spacing in both directions across the survey area.

\begin{figure}
	\centering
	\includegraphics[width=1.\linewidth]{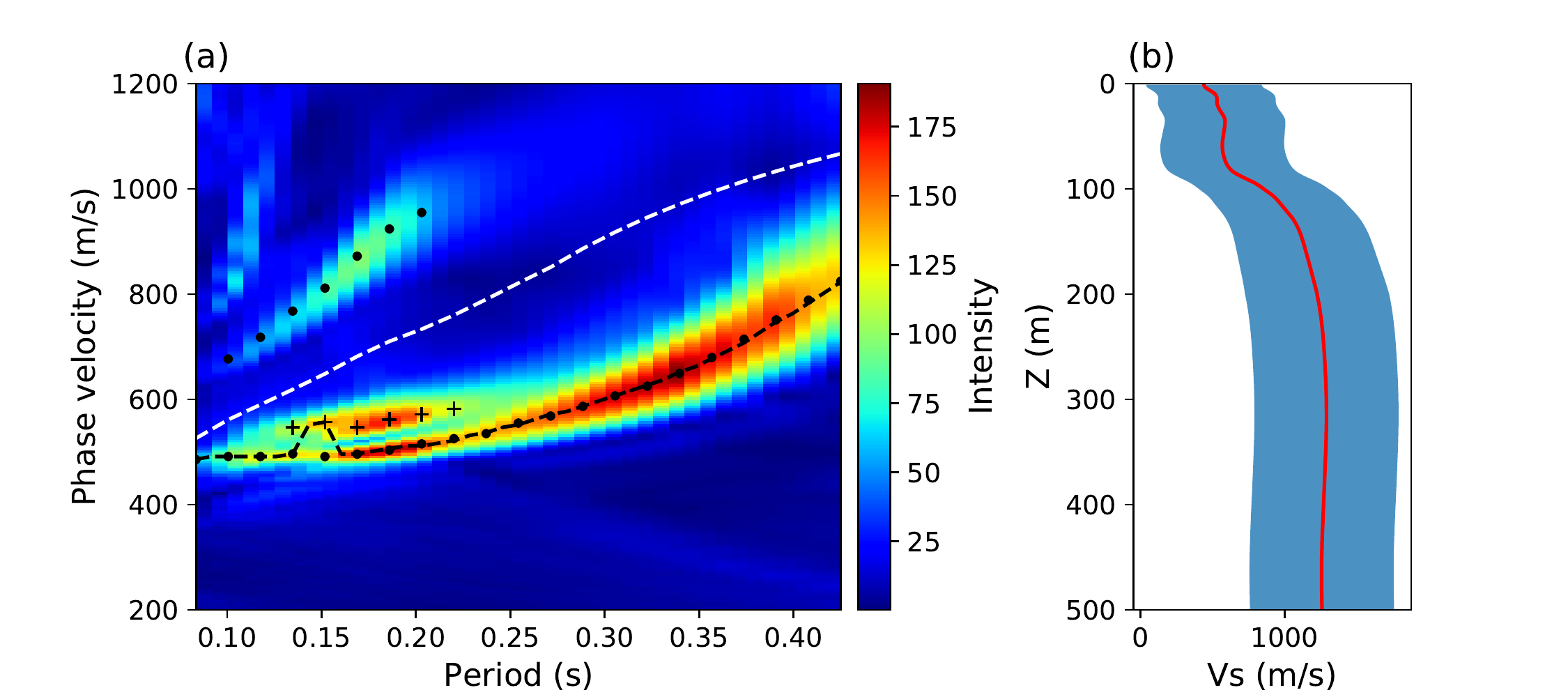}
	\caption{ \textbf{(a)} An example spectrum obtained using the f-c analysis at a specific location (red star in Figure \ref{fig:data}a). Black dots denote the picked phase velocities, black pluses show the phase velocities associated with the second branch for the fundamental mode and the black dashed line shows the phase velocity dispersion curve associated with the maximum energy at each period. The white line is used to separate the two modes. \textbf{(b)} shows the prior information, which is a Uniform distribution with an interval up to 1000 m/s at each depth. The red line shows the mean of the prior pdf.}
	\label{fig:dispersion}
\end{figure}

Figure \ref{fig:dispersion}a shows an example spectrum obtained using the above method (displayed as phase velocity versus period) at one specific geographic location (red star in Figure \ref{fig:data}a). The spectrum shows three modes. The phase velocity of the fundamental mode varies from 500 m/s to 800 m/s and contains two different branches at periods shorter than 0.25 s. This might be caused by effects such as multipathing and strong heterogeneity, or represents two different surface wave modes. To further understand this, we performed an inversion using one of the branches (black dots in Figure \ref{fig:dispersion}a) and modelled the first overtone using the obtained shear velocity profile (see Appendix A). The results show that the modelled first overtone is close to the mode with velocity higher than 600 m/s (see Figure A1). This therefore demonstrates that the two branches are probably not associated with different modes. The first overtone mainly appears in the period range from 0.1 s to 0.25 s with phase velocities varying from 620 m/s to 1000 m/s. The second overtone has much lower energy compared to the other two modes, and we discarded this mode in the inversion.

Traditionally for each mode those phase velocities associated with the peak energy are used as data to constrain the subsurface shear velocity. However, for those modes that have complex structures in the spectrum (such as the fundamental mode in Figure \ref{fig:dispersion}a), it becomes difficult to determine the correct phase velocity dispersion curve to use as these may have apparent jumps between neighbouring periods (e.g., the dashed black line in Figure \ref{fig:dispersion}a). One way to reduce this issue is to impose continuity or smoothness constrains on the dispersion curves. For example, in Figure \ref{fig:dispersion} the black dots are determined by forcing the dispersion curve to be smooth. Unfortunately the picked dispersion curve is then forced to follow one of the branches at shorter periods ($< 0.25 s$) which may not be the true dispersion curve generated by the subsurface, and consequently the inverted velocity structure may be biased. In addition, for large datasets the dispersion curves need to be determined automatically which introduces difficulties to balance higher energy against the smoothness of dispersion curves, and therefore may cause errors in the estimated phase velocities. In the next section, we propose a method which directly uses the f-c spectrum as data to constrain the subsurface velocity structure, thus avoiding these issues.

\section{Shear-wave velocity inversion}
The surface wave dispersion information obtained as above can be used to constrain the subsurface shear velocity structure, which involves solving a nonlinear and nonunique inverse problem. In this study we use Bayesian inference to characterize the fully nonlinear uncertainty of the solution.
\subsection{Bayesian inference}
In Bayesian inference one constructs a so-called posterior pdf $p(\mathbf{m}|\mathbf{d}_{\mathrm{obs}})$ of velocity model $\mathbf{m}$ given the observed data $\mathbf{d}_{\mathrm{obs}}$, by combining prior information with new information contained in the data. According to Bayes' theorem,
\begin{equation}
	p(\mathbf{m}|\mathbf{d}_{\mathrm{obs}}) = \frac{p(\mathbf{d}_{\mathrm{obs}}|\mathbf{m})p(\mathbf{m})}{p(\mathbf{d}_{\mathrm{obs}})}
	\label{eq:bayes}
\end{equation}
where $p(\mathbf{m})$ describes the prior information of model $\mathbf{m}$ that is independent of the current data, $p(\mathbf{d}_{\mathrm{obs}}|\mathbf{m})$ is called the \textit{likelihood} which describes the probability of observing $\mathbf{d}_{\mathrm{obs}}$ if model $\mathbf{m}$ was true, and $p(\mathbf{d}_{\mathrm{obs}})$ is a normalization factor called the \textit{evidence}.

The prior information is critical for Bayesian inference. To construct a more informative prior distribution than the commonly-used Uniform distribution with little or no depth dependence, we first conduct a set of inversions at multiple geographic locations using a Uniform prior pdf from 300 m/s to 1500 m/s, which spans the range of shear velocities in the upper 500 m according to a variety of similar studies \citep{lee2008integrated, mordret2014ambient, chmiel2019ambient, zhang20201d}. The average of the mean models from these inversions are then used as the mean of the prior pdf, and we construct a Uniform distribution with a width of 1000 m/s (larger than four standard deviations obtained from the previous inversions) at each depth (Figure \ref{fig:dispersion}b). This prior information improves the depth resolution and constrains the subsurface structure better than an identical Uniform distribution across the depth ranges \citep{yuan2018probabilistic}.

\subsection{Energy likelihood function}
In the f-c spectrum higher intensity (energy) means higher probability that the associated phase velocity represents the true phase velocity of a surface wave mode. Based on this assumption, we can directly use the spectrum as data and write a new likelihood function. Define $\mathbf{E}$ as the matrix representing the f-c spectrum and $E(T,c)$ as the energy at period $T=1/f$ and phase velocity $c$, and assuming that the energy at each value of T and c has exponentially decaying probability away from the maximum energy value at that period, the likelihood can be expressed as:
\begin{equation}
	p(\mathbf{E}|\mathbf{m}) = \frac{1}{Z} \mathrm{exp} [- \sum_{i} \frac{max(\mathbf{E}(T_{i},\cdot))-E(T_{i},c_{i}(\mathbf{m}))}{\sigma_{i}^{2}}]
	\label{eq:energy_likelihood} 
\end{equation}
where $T_{i}$ is the $i^{th}$ period, $max(\mathbf{E}(T_{i},\cdot))$ is the maximum energy at period $T_{i}$ which guarantees that the exponent is negative, $c_{i}(\mathbf{m})$ is the phase velocity at period $T_{i}$ predicted using model $\mathbf{m}$, $\sigma_{i}$ is a scaling factor and $Z$ is a normalization factor. The scaling factor $\sigma_{i}$ is generally unknown, so we treat it as an additional parameter and estimate it hierarchically \citep{malinverno2004expanded}. 

For multimodal inversion the energy of the fundamental mode may dominate the likelihood function in equation \ref{eq:energy_likelihood} (for example, see Figure \ref{fig:dispersion}a), and consequently the inverted results can be biased because models may have apparently larger likelihood values if their predicted higher modes also fit the fundamental mode energy. We therefore separate each mode by windowing out other modes. For example, define $\mathbf{E}_{j}$ to be the spectrum of the $j^{th}$ mode after other modes have been windowed out. Then the likelihood function becomes:
\begin{equation}
	p(\mathbf{E}|\mathbf{m}) = \frac{1}{Z} \mathrm{exp} [- \sum_{ij} \frac{max(\mathbf{E}_{j}(T_{i},\cdot))-E_{j}(T_{i},c_{i}(\mathbf{m}))}{\sigma_{ij}^{2}}]
	\label{eq:energy_likelihood2} 
\end{equation}
Although this requires that we define a window function for each mode, this process is usually straightforward. For example, the white dashed line in Figure \ref{fig:dispersion}a shows the boundary used to separate the first two modes; this same line is used for all other spectra across the survey area since it appeared appropriate for a large number of spectra examined manually.

\subsection{Reversible-jump Markov chain Monte Carlo (rj-McMC)}
We use reversible-jump Markov chain Monte Carlo (rj-McMC) to generate samples from the posterior pdf. The rj-McMC method is a generalised version of the Metropolis-Hastings algorithm \citep{metropolis1949monte, hastings1970monte}, which allows the number of model parameters to be variable in the inversion \citep{green1995reversible}. Thus the parameterization of the seismic velocity model can itself be determined by the data and prior information. The method has been applied in a range of geophysical applications \citep{malinverno2002parsimonious, bodin2009seismic, young2013transdimensional, galetti2015uncertainty, piana2015local, galetti2017transdimensional, zhang20183, zhang20201d, zhang2020imaging}. In this study we use the method to solve the surface wave dispersion inversion problem.

In rj-McMC one constructs a (Markov) chain of samples by perturbing the current model $\mathbf{m}$ using a proposal distribution $q(\mathbf{m}'|\mathbf{m})$ to generate a new model $\mathbf{m}'$, and by accepting or rejecting this new model with a probability $\alpha(\mathbf{m}'|\mathbf{m})$ called the acceptance ratio:
\begin{equation}
	\alpha(\mathbf{m}'|\mathbf{m}) = \mathrm{min} [1, \frac{p(\mathbf{m}')}{p(\mathbf{m})} \times 
	\frac{q(\mathbf{m}|\mathbf{m}')}{q(\mathbf{m}'|\mathbf{m})} \times
	\frac{p(\mathbf{d}_{\mathrm{obs}}|\mathbf{m}')}{p(\mathbf{d}_{\mathrm{obs}}|\mathbf{m})} \times
	|\mathbf{J}|]
	\label{eq:acceptance_ratio}
\end{equation}
where $\mathbf{J}$ is the Jacobian matrix of transforming $\mathbf{m}$ to $\mathbf{m}'$ and is used to account for any volume changes of parameter space during jumps between dimensionalities. In our case the Jacobian matrix is an identity matrix \citep{bodin2009seismic}.

For surface wave dispersion inversion beneath each geographical location we use a set of layers to parameterize the subsurface, which can be changed in different ways within the rj-McMC algorithm: adding a new layer, removing a layer, changing layer positions and changing layer velocities. There is also another perturbation related to the hypeparameters of the likelihood function: changing the scaling factor $\sigma_{ij}$ in equation \ref{eq:energy_likelihood2} or standard deviation in the Gaussian likelihood function. After each perturbation of the current model $\mathbf{m}$, the acceptance ratio $\alpha$ is computed using equation \ref{eq:acceptance_ratio} and is compared with a random number $\gamma$ generated from the Uniform distribution on [0,1]. If $\gamma < \alpha$ the new model is accepted; otherwise the new model is rejected and the current model is repeated as a new sample in the chain. This process guarantees that the generated samples are distributed according to the posterior pdf if the number of samples tends to infinity \citep{green1995reversible}.

\subsection{A 1D example}

\begin{figure}
	\includegraphics[width=1.\linewidth]{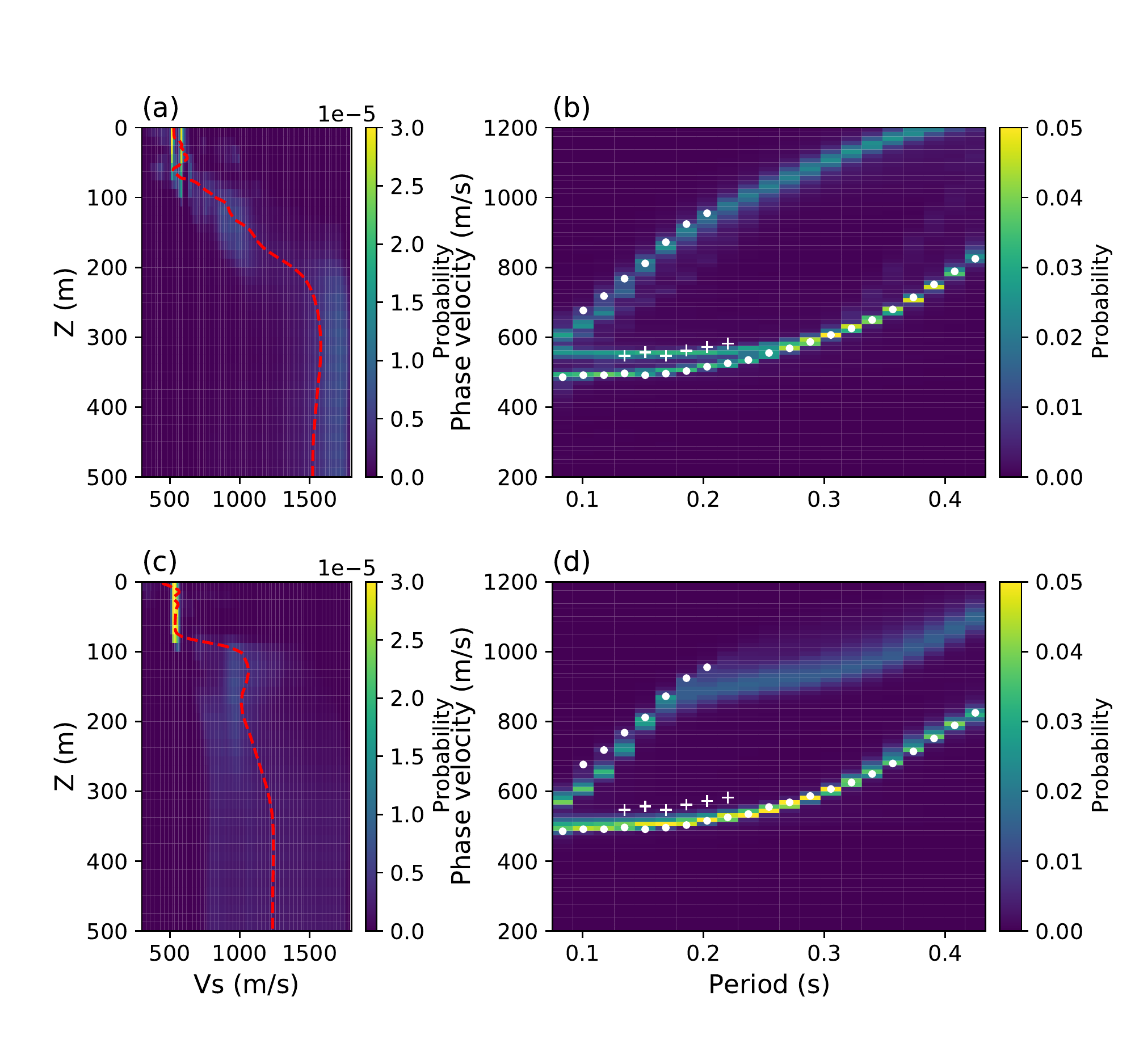}
	\caption{\textbf{(a)} The shear velocity and \textbf{(b)} the phase velocity marginal posterior distributions obtained using the posterior samples of the energy likelihood function for the spectrum in Figure \ref{fig:dispersion}a. The phase velocity distribution is obtained by modelling the phase velocity for all posterior samples. \textbf{(c)} and \textbf{(d)} show the associated shear velocity and phase velocity marginal distributions obtained by the traditional method of using the picked phase velocities shown by black dots in Figure \ref{fig:dispersion}a. Red dashed lines show the mean velocity profile. White dots and pluses are associated with the black dots and pluses in Figure \ref{fig:dispersion} respectively.}
	\label{fig:example}
\end{figure}

We first apply the above method to the dispersion data in Figure \ref{fig:dispersion}a using both the energy likelihood function and the Gaussian likelihood function and compare their results. We use both the fundamental mode and the first overtone with 21 equally spaced periods from 0.0835 s to 0.425 s. The prior pdf of shear velocity is shown in Figure \ref{fig:dispersion}b and the prior pdf of the number of layers is set to be a discrete Uniform distribution between 2 and 25. For the energy likelihood function the prior pdf of the scaling factor $\sigma_{ij}$ is chosen to be a Uniform distribution between 0.5 and 20, and for the Gaussian likelihood function in the traditional method the prior pdf of the data noise (standard deviation of the Gaussian distribution) is a Uniform distribution between 0.05 and 50. We use Gaussian distributions as the proposal distribution: the width of the Gaussian distribution for the fixed-dimensional steps (changing layer positions, changing layer velocities and changing likelihood hyperparameters) is chosen by trial and error to ensure an acceptance ratio between 20 and 50 percent; the width of the transdimensional steps (adding or removing a layer) is selected to produce the maximum acceptance ratio. For each inversion we run six chains, and each chain contains 800,000 iterations with a burn-in period of 300,000 during which all samples are discarded. After the iteration each chain is thinned (decimated) by a factor of 50, and the remaining samples are used for the subsequent inference.
 
Figure \ref{fig:example}a and \ref{fig:example}c show the marginal distributions of shear velocity obtained using the energy likelihood function and the Gaussian likelihood function respectively. The corresponding phase velocity distributions generated by those posterior samples are shown in Figure \ref{fig:example}b and \ref{fig:example}d respectively. The shear velocity marginal distribution obtained using the energy likelihood function shows clearly multimodal distributions in the near surface ($< 100$ m), which are associated with the two branches in the f-c spectrum (Figure \ref{fig:dispersion}a and Figure \ref{fig:example}b). In comparison the marginal distribution obtained using the Gaussian likelihood function shows a unimodal distribution and the predicted data only fit the single branch on which the phase velocities were picked. Since we do not know which branch reflects the most appropriate phase velocity a priori, the shear velocity obtained using the Gaussian likelihood function is biased and the estimated uncertainty failed to take account of the full, multi-modal uncertainty in the data. In contrast, by directly using the spectrum as data one can embed all data uncertainty in the likelihood function and therefore obtain a less biased result. In addition, the phase velocity distribution of the first overtone obtained using the energy likelihood function (Figure \ref{fig:example}b) fits the picked phase velocity better than that obtained using the Gaussian likelihood function. This suggests that there is inconsistency between the picked phase velocities of the fundamental mode and the first overtone, and therefore further demonstrates the necessity of including the second branch in the likelihood function. In the deeper part ($>100$ m) the velocity obtained using the energy likelihood function increases from 600 m/s at 100 m depth to 1500 m/s around 220 m and stays almost constant down to 500 m; whereas the velocity obtained using the Gaussian likelihood function is around 1200 m/s across the whole depth from 100 m to 500 m. This is probably because the picked phase velocities (black dots in Figure \ref{fig:dispersion}a) are not sensitive to the deeper part ($> 200$m), and consequently the shear velocity in the deeper part are dominated by the prior pdf. In comparison the energy likelihood function uses all the information contained in the spectrum and constrains deeper structure better. For example, at periods longer than 0.22s phase velocities are not determined for the first overtone because of its lower energy, whereas the information is still used in the energy likelihood function to constrain deeper structures. This can also be observed in the phase velocity distributions predicted from posterior samples (Figure \ref{fig:example}b and d), where the first overtone distribution obtained using the energy likelihood function at longer periods ($>0.22$ s) is more similar to the spectrum than that obtained using the Gaussian likelihood function. Thus, by directly using the spectrum as data the new energy likelihood function can use more information in the data and can obtain more accurate results than the traditional method.

\section{3D results}

\begin{figure}
	\includegraphics[width=1.\linewidth]{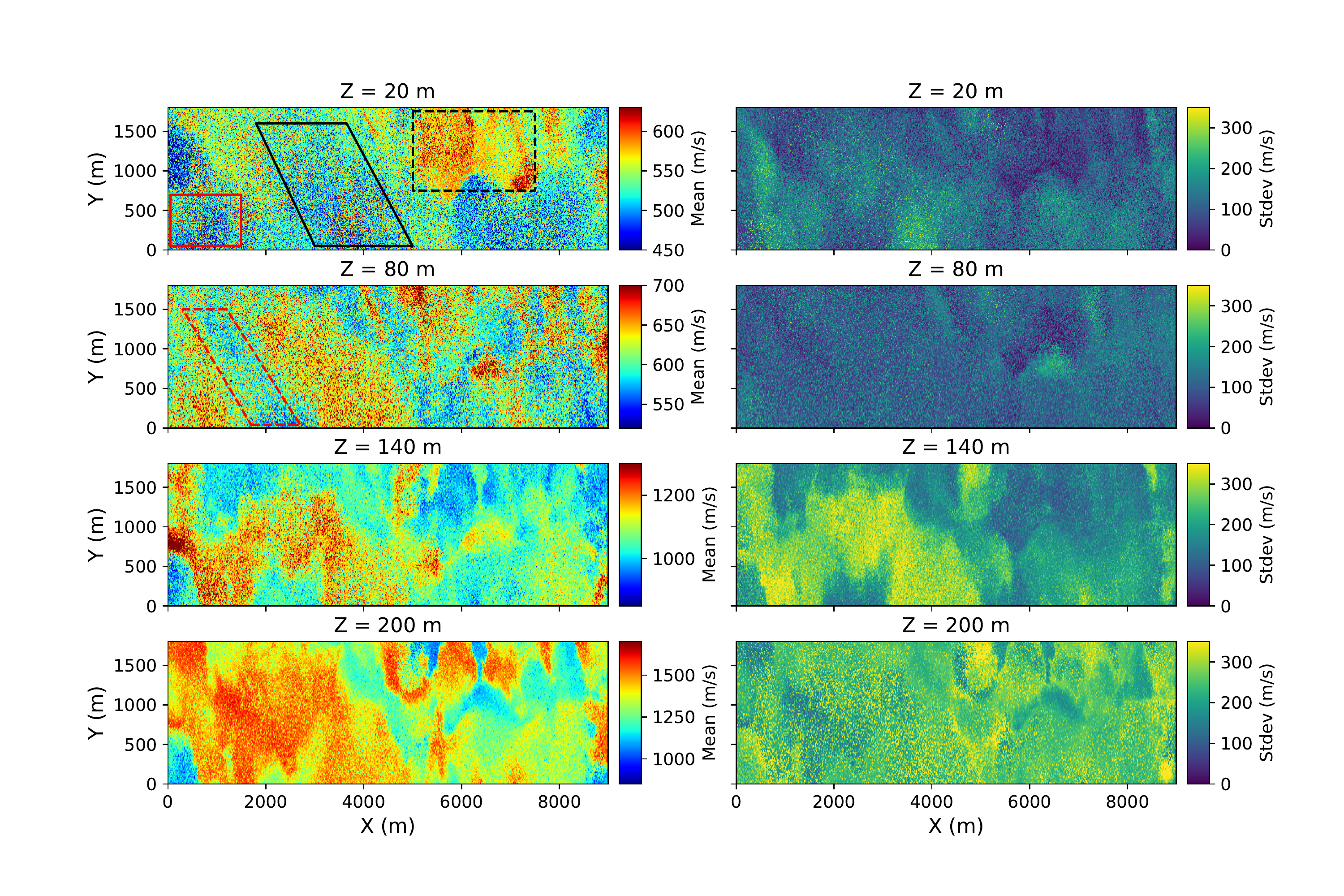}
	\caption{The mean (\textbf{left panel}) and standard deviation (\textbf{right panel)} of shear velocities obtained using the energy likelihood across horizontal slices at depths 20 m, 80 m, 140 m and 200 m. Boxes highlight velocity anomalies which are referred to in the text.}
	\label{fig:energy_results}
\end{figure}

\begin{figure}
	\includegraphics[width=1.\linewidth]{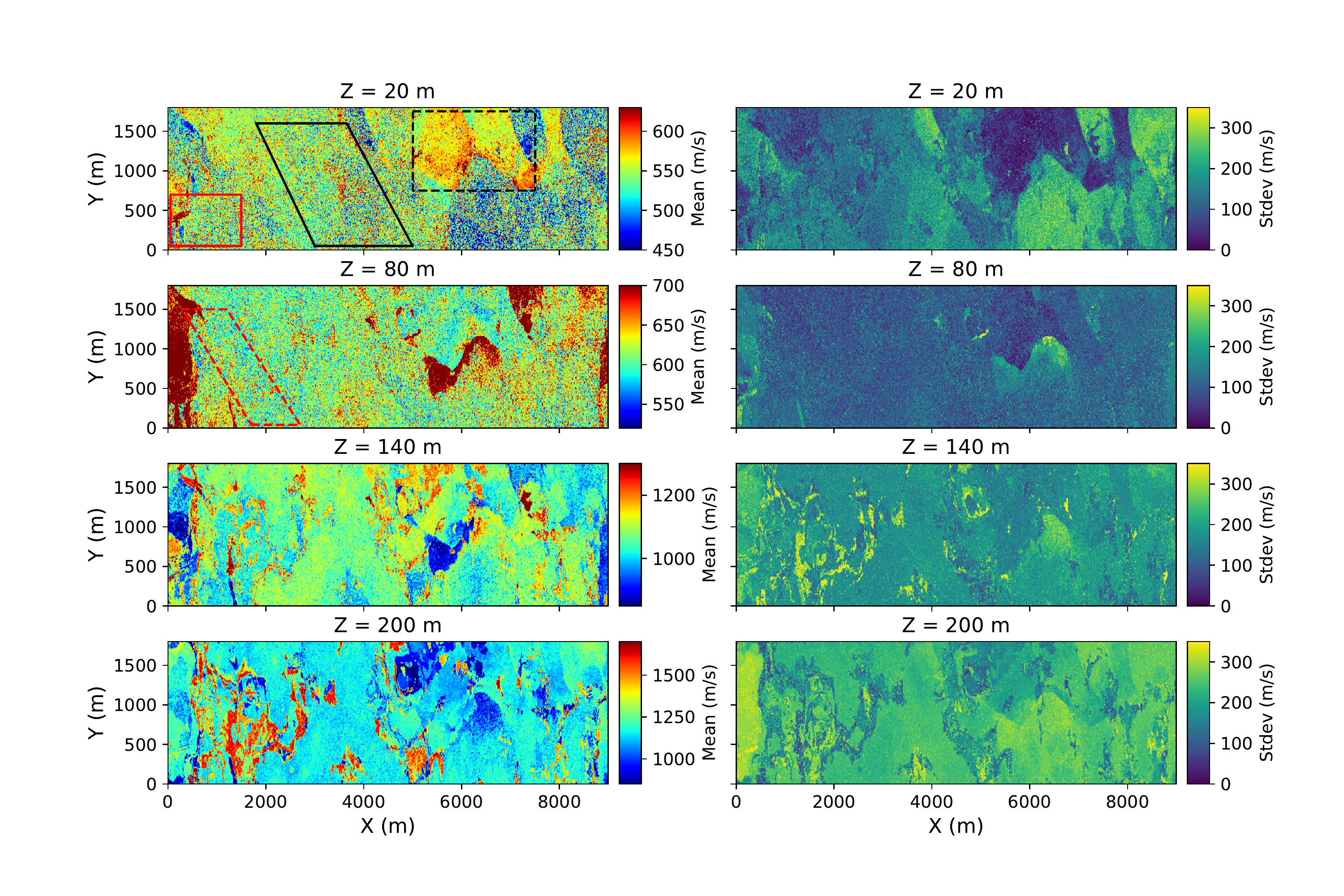}
	\caption{The mean (\textbf{left panel}) and standard deviation (\textbf{right panel)} of shear velocities obtained using the traditional method with picked phase velocities across horizontal slices at depths 20 m, 80 m, 140 m and 200 m. Boxes are the same as in Figure \ref{fig:energy_results}.}
	\label{fig:picks_results}
\end{figure}

\begin{figure}
	\includegraphics[width=.8\linewidth]{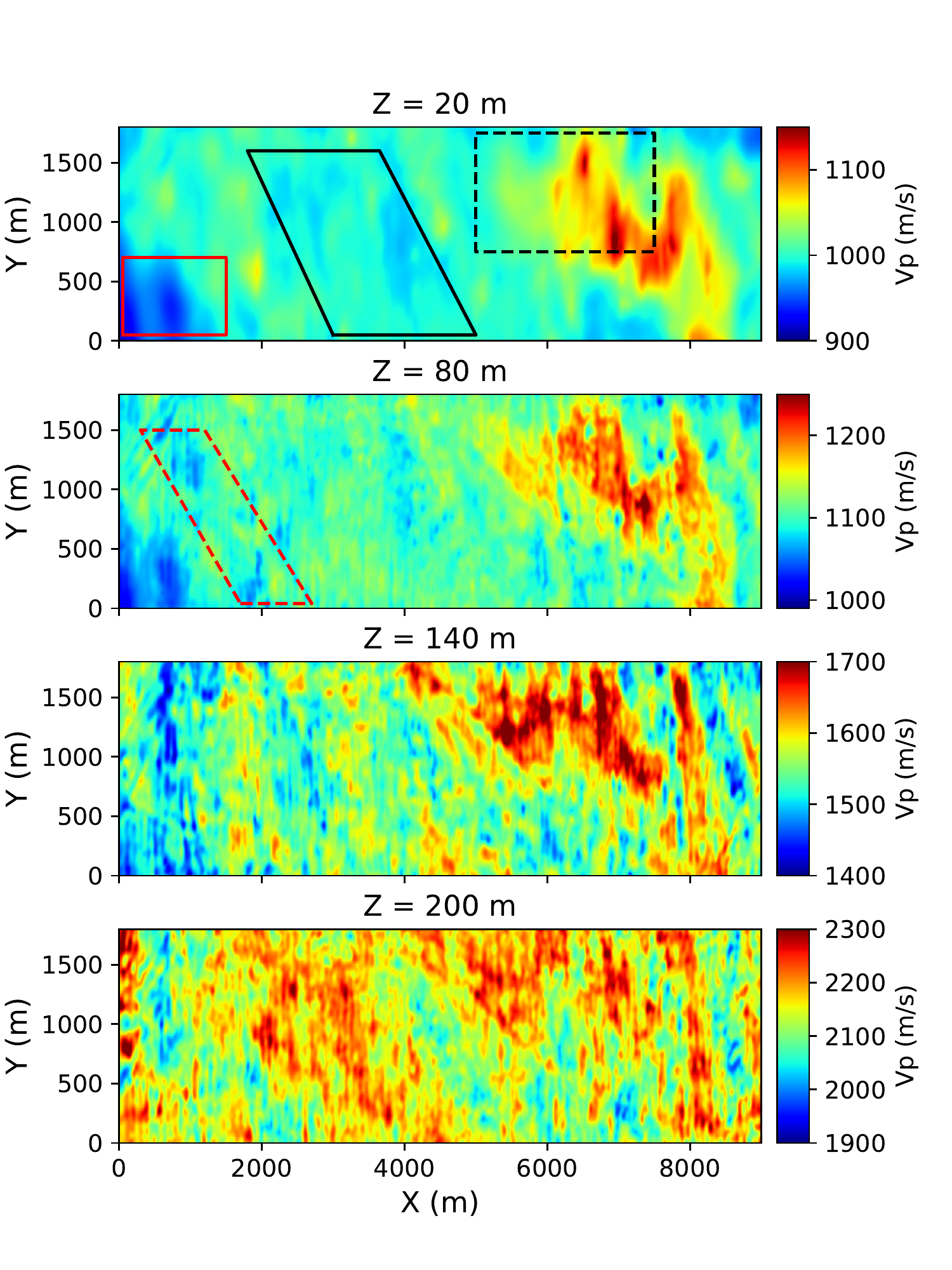}
	\caption{The P-wave velocity model obtained using refraction tomography across horizontal slices at depths 20 m, 80 m, 140 m and 200 m. Boxes are the same as in Figure \ref{fig:energy_results}.}
	\label{fig:vp_results}
\end{figure}

\begin{figure}
	\includegraphics[width=1.\linewidth]{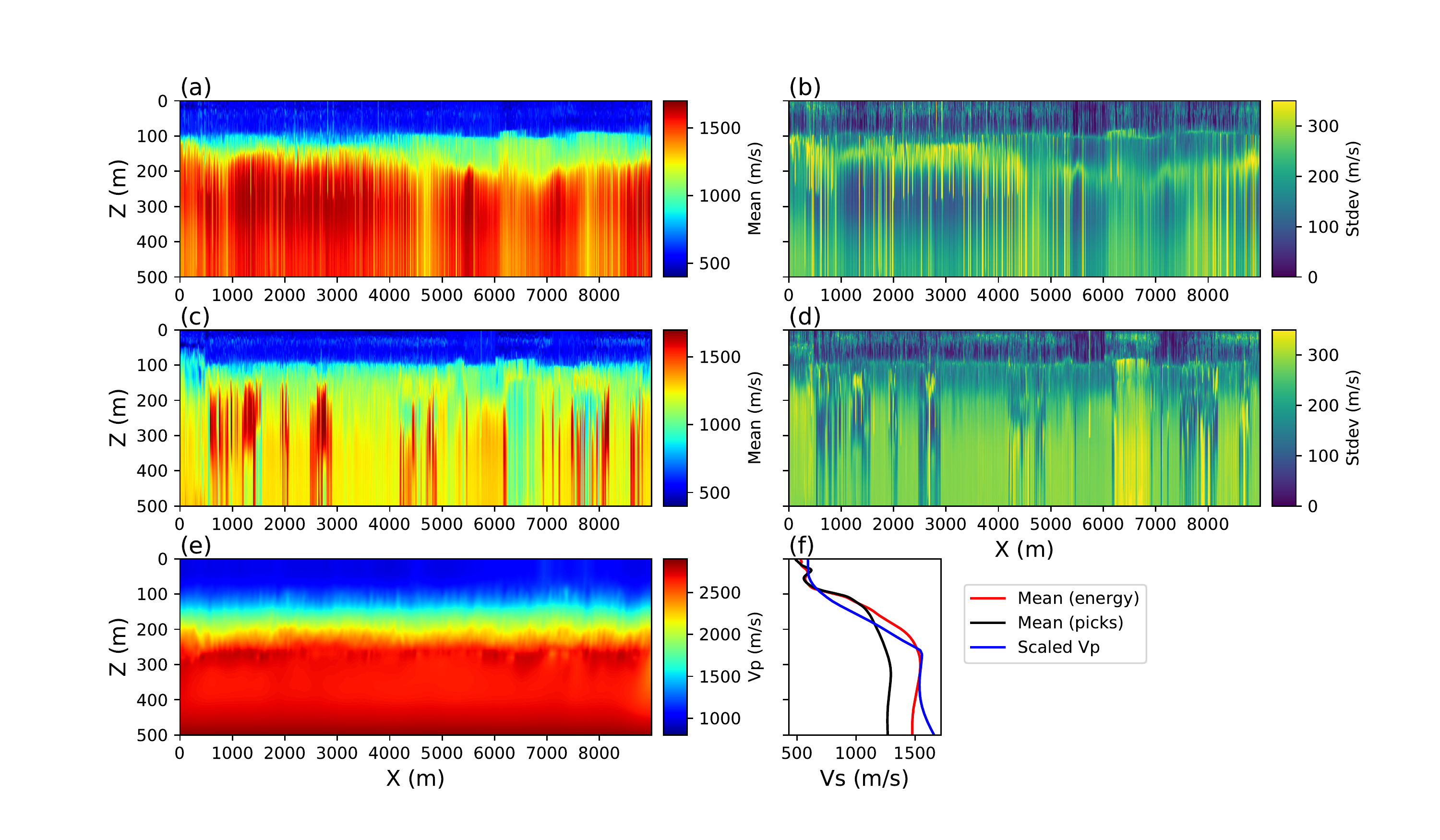}
	\caption{The mean and standard deviation model obtained using the energy likelihood function (\textbf{a} and \textbf{b}) and the traditional method (\textbf{c} and \textbf{d}) along a vertical section at Y = 875 m. \textbf{(e)} shows the vertical section of the P-wave velocity model obtained using refraction tomography. \textbf{(f)} shows the average shear velocity profiles across the section obtained using the suite of methods. The P-wave velocity is converted to shear velocity using a Vp/Vs ratio of 1.73.}
	\label{fig:vertical_slice}
\end{figure}

To obtain 3D shear velocity models we perform 1D inversions to all of the spectra across the survey area. For each inversion at each geographic location the inversion is conducted in the same way as described in the previous section with the same prior pdf and proposal pdf. To compare the results we use both the energy likelihood function and the Gaussian likelihood function around picked phase velocities. For each spectrum we automatically determine the phase velocity from longer periods to shorter periods: at each period the phase velocity is determined as the local maximum energy point whose phase velocity is closest to the already picked phase velocity at the neighbouring period, such that the picked dispersion curve is as continuous as possible. To ensure the quality of picked phase velocities, we only determine the phase velocity at frequencies with sufficiently high energy, such that the picked energy is at least 1.8 times higher than the average energy at each frequency.

Figure \ref{fig:energy_results} shows the mean and standard deviation models at the depth of 20 m, 80 m, 140 m and 200 m obtained using the energy likelihood function. Among various structures we have highlighted several velocity anomalies using boxes which we will refer to below. Overall the standard deviation model shows lower uncertainty at the shallower part (20 m and 80 m) and higher uncertainty at the deeper part (140 m and 200 m) due to the fact that seismic surface waves are more sensitive to near surface structure. At all the depths the standard deviation shows similar features to the mean model, but notice that at 20 m depth higher velocity anomalies are associated with lower uncertainties, whereas at 140 m depth higher velocity anomalies correspond to higher uncertainties. This phenomenon has also been observed previously \citep{zhang2020seismic, zhang2020variational, zhang2021bayesiana}, and the different correlation between velocity anomalies and uncertainties at different depths suggests that there is a complex, nonlinear relationship between seismic velocity and phase velocity data.   

Figure \ref{fig:picks_results} shows the mean and standard deviation models obtained using the Gaussian likelihood function at the same depths. Overall the mean model obtained using the Gaussian likelihood function shows more small scale variations than that obtained using the energy likelihood function. This is probably caused by errors in the picked phase velocities, which is inevitable when the phase velocities are estimated automatically. At 20 m depth the mean model shows similar structures to those observed in the previous results in Figure \ref{fig:energy_results}, for example the northern high velocity anomaly denoted by black dashed boxes and the low velocity anomalies in the sourtheast and northwest. However, the low velocity anomalies that are denoted by black and red solid line boxes in Figure \ref{fig:energy_results} are not present in Figure \ref{fig:picks_results}. Similar to the previous results, at 80 m depth there are small scale variations in the east, but the low velocity anomaly denoted by red dashed line box and the high velocity anomaly next to this low anomaly are not visible. At greater depths (140 m and 200 m) the mean velocity model is significantly different from the previous results and contains many small scale structures which probably do not reflect the true geologic structure as the scale of these structures is smaller than the scale (300 m) used for stacking which implicitly imposes smoothness to the velocity structure. This suggests that the traditional method can cause bias in the inverted velocity structure because of errors in the picked phase velocity and loss of useful information in the data. Similarly to the previous results, the standard deviation model shows similar features to the mean model, and the uncertainty is lower in shallower parts.

To further understand the results we compare the above shear velocity models with P-wave velocity model (Figure \ref{fig:vp_results}) obtained using refraction tomography from the same dataset \citep{buriola2021ultra}. At 20 m depth the Vp model shows similar features to the shear velocity models. For example, there is a high velocity anomaly in the east between X = 6000 m and 8000 m, which may be related to the high velocity anomaly (black dashed line box) observed in the shear velocity models even though they are not at exactly the same location. Similarly to the shear velocity model obtained using the energy likelihood function, there is a low velocity anomaly between X = 2000 m and 5000 m in the southeastern direction (black solid line box) and a low velocity anomaly in the southwestern corner (red solid line box). This strongly suggests that the new energy likelihood method is effective since those low velocity anomalies are not visible in the results obtained using the traditional method (Figure \ref{fig:picks_results}). Even though at depths of 80 m and 140 m the Vp model is different from both shear velocity models, there are still similarities between the Vp model and the shear velocity model obtained using the energy likelihood function. For example, at 80 m depth there is a low velocity channel in the southeastern direction (denoted by red dashed line boxes) in the west of both models. At 200 m depth the Vp model shows more similarities to the shear velocity model obtained using the energy likelihood function: both models show a high velocity anomaly between X = 2000 m and 5000 m in the southeastern direction across the area (the same location as denoted by the black solid line box at 20 m depth) and a low velocity channel to the east of this high velocity anomaly. In the east (X $>$ 4000 m) there are similar small scale high velocity anomalies in both models. In comparison the shear velocity model obtained using the traditional method is very different from the Vp model. This again shows that the new energy likelihood function generates more accurate shear velocity models than the traditional method.

Figure \ref{fig:vertical_slice} shows vertical sections through the different models at Y = 900 m. To further compare these models, in Figure \ref{fig:vertical_slice}f we show the average velocity profile across the section of each model with the P-wave velocity converted to shear velocity using a factor of 1.73 \citep{pickett1963acoustic}. In the near surface ($< 100 $m) all three models show a lower velocity layer (Figure \ref{fig:vertical_slice}f), although the two shear velocity models show more complex structures than the P-wave velocity. Between depths of 100 m and 230 m the shear velocity obtained using the energy likelihood function increases from 600 m/s to 1500 m/s which is consistent with the trend of P-wave velocity, and both models show a high velocity layer below 200 m. In comparison the shear velocity obtained using the traditional method is around 1300 m/s below 100 m. A similar phenomenon was observed in section 3 in the 1D example, which is caused by the fact that the picked phase velocities are not sensitive to the deeper structure. For example, the standard deviation model obtained using the traditional method shows higher uncertainties ($> 300 $ m/s) below 200 m, whereas that obtained using the energy likelihood function shows a much lower uncertainty ($< 100$ m/s) between 200 m and 400 m. In addition, the shear velocity model obtained using the traditional method shows higher lateral spatial variations in deeper parts of the section ($> 200$ m) than the other two models. This further demonstrates the importance of using all information contained in the spectrum, rather than using only picked phase velocities. Note that both shear velocity standard deviation models show higher uncertainties at the layer boundary, which has been observed previously \citep{galetti2015uncertainty, zhang20183} and reflects the uncertainty of boundary positions. 


\section{Discussion} 
The McMC method is generally computationally expensive. For the above inversion each chain took 0.173 CPU hours using one core of an Intel Xeon CPU, which results in a total of 1.04 CPU hours for each geographic location and 107,015.4 CPU hours for all 102,817 inversions across the survey area. However, the inversions are fully parallelizable since the inversion at each geographical location is independent of other inversions. For example, the above inversion across the whole survey area took 15 hours using 7,200 CPU cores. We also note that other methods can be used to further improve the computational efficiency, for example, Hamiltonian Monte Carlo \citep{duane1987hybrid,fichtner2018hamiltonian}, Langevin Monte Carlo \citep{roberts1996exponential, siahkoohi2020uncertainty}, variational inference \citep{nawaz2018variational, zhang2020seismic, zhang2020variational, zhao2021bayesian, zhang2021bayesiana} and neural network inversion \citep{meier2007aglobal,meier2007bglobal,earp2020probabilistic,zhang2021bayesianb}. 

In the above results the phase velocities used for the traditional method are determined automatically from the f-c spectrum, and these may contain errors due to the complexity of the f-c spectrum. We therefore note that the results obtained using the traditional method may be improved if the phase velocities can be determined more accurately. However, this usually requires deliberately and manually picking of a dispersion curve for each spectrum, which restricts its application to a relatively small number of spectra. By contrast the new energy likelihood function directly using the spectrum as data and can easily be applied to larger datasets. 

We performed the surface wave dispersion inversion independently at each geographic location, which loses lateral spatial correlations between neighbouring velocities and can cause biases in the final results \citep{zhang20183}. For example, there are horizontal discontinuities in the shear velocity mean and standard deviation models (see Figure \ref{fig:vertical_slice}), which probably do not reflect the real subsurface structure. To include the lateral spatial correlation and to further improve the results a 3D parameterization may be used instead of the 1D parameterization, for example, 3D Voronoi tessellation can be used in surface wave tomography \citep{zhang20183, zhang20201d}.

In this study we used the spectrum obtained by stacking wavefields in the f-c domain, which requires a dense receiver array. In cases where only a sparse array is available, the energy likelihood can still be used to perform the inversion. For example, one could directly use the spectrum obtained between each source-receiver or receiver-receiver as the data to conduct phase or group velocity tomography using earthquake-generated surface waves or ambient noise data.
     
\section{Conclusion}
In this study we introduced a new likelihood function for seismic surface wave dispersion inversion, called the energy likelihood function which directly uses the spectrum as data. We applied the new likelihood function to image the subsurface shear velocity structure using surface wave data recorded by a dense array, and compared the results with those obtained using the traditional method. The results showed that the new likelihood function can take account of all information contained in the spectrum and produce a less biased result than that obtained using the traditional method. In addition the velocity model obtained using the new likelihood function is more similar to the P-wave velocity structure than that obtained using the traditional method. Because the new likelihood function directly uses the spectrum as data, it requires less effort to apply to large datasets than the traditional method, since the latter requires that we determine the phase velocity for each spectrum either manually or automatically and cannot be applied easily to large datasets. We thus conclude that the new energy likelihood function provides a powerful way to conduct surface wave dispersion inversion.

\section*{Acknowledgments}
The authors thank the Edinburgh Imaging Project sponsors (BP and Total) for supporting this research. We would like thank BP and ADNOC for providing the seismic data and CGG for the refraction tomography model.

\bibliographystyle{plainnat}
\bibliography{bibliography}

\appendix
\section{mode validation}
\begin{figure}
	\centering
	\includegraphics[width=1.\linewidth]{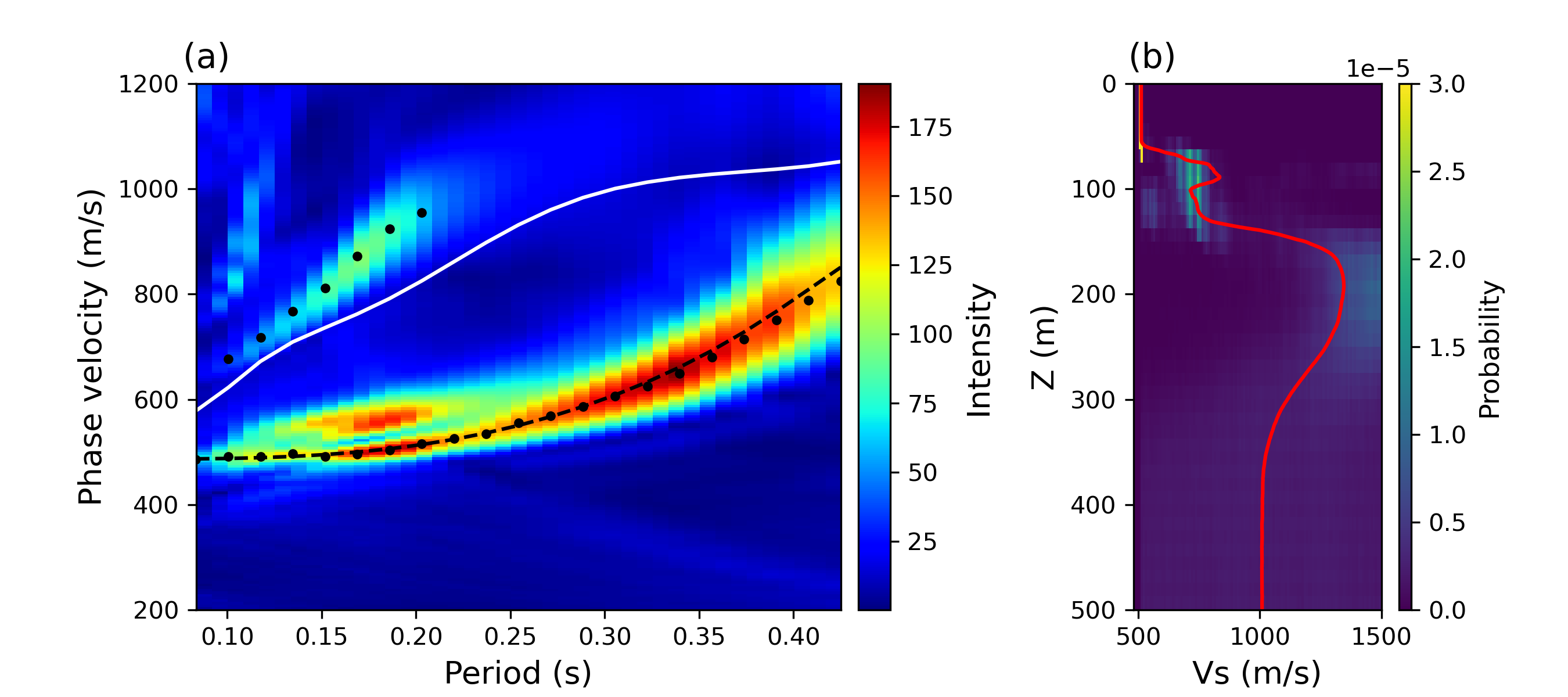}
	\caption{\textbf{(a)} The spectrum obtained using f-c analysis for the location in Figure \ref{fig:data}b (red star). Black dots show the picked phase velocities. \textbf{(b)} The marginal posterior distribution of shear velocities obtained using only the picked phase velocities of the fundamental mode. The red line shows the mean shear velocity profile. The black dashed line and the white line in \text{(a)} show the predicted phase velocities for the fundamental mode and the first overtone respectively, modelled using the posterior mean model.}
	\label{fig:mode_validation}
\end{figure}

To understand the two branches that appear at short periods ($< $0.25 s) in the f-c spectrum in Figure \ref{fig:dispersion}, and in particular to test whether the two branches represent two different modes, we first perform an inversion using the rj-McMC algorithm using one of the branches as data (black dots in Figure \ref{fig:mode_validation}). The prior pdf of the shear velocity is set to be a Uniform distribution between 300 m/s and 1500 m/s. For the likelihood function we use the traditional Gaussian distribution. The inversion is then conducted in the same way as described in section 3.4 with the same prior pdf for the number of layers and noise hyperparameters and the same proposal pdf. Figure \ref{fig:mode_validation}b shows the obtained mean and the marginal distribution of the shear velocity. We then use the mean shear velocity profile to simulate phase velocities of the fundamental mode and the first overtone. While the modelled fundament mode phase velocity (black dashed line in Figure \ref{fig:mode_validation}a) fits the data used, the modelled phase velocity of the first overtone (while line in Figure \ref{fig:mode_validation}a) is significantly closer to the mode with velocity higher than 600 m/s than to the other branch appearing in the fundamental mode. This clearly demonstrates that the two branches are unlikely to represent two different modes, and instead represent an effect such as the multipathing of the seismic energy of the fundamental model or the strong lateral heterogeneity. 
\label{lastpage}

\end{document}